\documentclass[12pt]{iopart}

\usepackage{graphicx}
\usepackage{epstopdf}

   \AtEndOfPackage{%
      \g@addto@macro\Gin@extensions{.eps.gz,.eps.zip}%
      \@namedef{Gin@rule@.eps.gz}#1{{pdf}{.pdf}{`gunzip -c #1 | epstopdf -f >\Gin@base.pdf}}%
      \@namedef{Gin@rule@.eps.zip}#1{{pdf}{.pdf}{`unzip -p #1 | epstopdf -f >\Gin@base.pdf}}%
   }%

\usepackage{cite}
\usepackage{color}
\usepackage{multirow}

\usepackage{epstopdf} 

\begin{document}

\setlength{\parindent}{0pt}

\title[ ]{Effects of fast atoms and energy-dependent secondary electron emission yields in PIC/MCC simulations of capacitively coupled plasmas}

\author{A. Derzsi$^1$, I. Korolov$^1$, E. Sch\"{u}ngel$^2$,  Z. Donk\'o$^1$, J. Schulze$^2$}

\address{$^1$Institute for Solid State Physics and Optics, Wigner Research Centre for Physics, Hungarian Academy of Sciences, 1121 Budapest, Konkoly Thege Mikl\'os str. 29-33, Hungary.\\ $^2$Department of Physics, West Virginia University, Morgantown, WV 26506, USA}
\ead{derzsi.aranka@wigner.mta.hu}

\begin{abstract}
In most PIC/MCC simulations of radio frequency capacitively coupled plasmas (CCPs) several simplifications are commonly made: (i) fast neutrals are not traced, (ii) heavy particle induced excitation and ionization are neglected, (iii) secondary electron emission from boundary surfaces due to neutral particle impact is not taken into account, and (iv) the secondary electron emission coefficient is assumed to be constant, i.e., independent of the incident particle energy and the surface conditions. Here, we examine the validity of these simplifications under conditions typical for plasma processing applications. We study the effects of including fast neutrals and using realistic energy-dependent secondary electron emission coefficients for ions and fast neutrals in simulations of CCPs operated in argon at 13.56 MHz and at neutral gas pressures between 5 Pa and 100 Pa. We find an increase of the plasma density and the ion flux to the electrodes under most conditions when heavy particles are included realistically in the simulation. The sheath widths are found to be smaller and the simulations are found to diverge at high pressures for high voltage amplitudes in qualitative agreement with experimental findings. By switching individual processes on and off in the simulations we identify their individual effects on the ionization dynamics and plasma parameters. While the gas-phase effects of heavy particle processes are found to be moderate at most conditions, the self-consistent calculation of the effective secondary electron yield proves to be important in simulations of CCPs in order to yield realistic results.

\end{abstract}


\submitto{\PSST}

\section{Introduction}

Low pressure capacitively coupled plasmas (CCP) are widely used for plasma processing: they are basic tools in applications such as plasma-enhanced chemical vapor deposition and plasma etching in the semiconductor industry, as well as in applications aimed at surface treatment in bio-engineering and medicine \cite{Lieberman_Book,Makabe_Book,Chabert_Book}. Their manifold applications, as well as their complex physics have been motivating extensive research in this field via modern experimental methods, analytical modeling, and computer simulation techniques.

The Particle-in-Cell (PIC) approach \cite{Birdsall_Book,Hockney_Book} combined with Monte Carlo (MC) type treatment of collision processes (known as PIC/MCC \cite{Birdsall_1991}) has become the prevailing self-consistent numerical method for the kinetic description of low-pressure CCPs \cite{Longo,Schneider,Verb}. In this approach ``superparticles'', representing a large number of real plasma particles, are traced, and their interaction is handled via the electric field calculated at points of a computational grid. This efficient simulation technique makes it possible to follow the spatio-temporal evolution of discharge characteristics and to obtain information about various plasma parameters, e.g. densities and fluxes of different plasma species, particle heating rates, rates of different collision processes, etc. 

In most of the PIC/MCC studies of CCPs the plasma species that are included in the models are the electrons and the ions formed from  the background gas. For instance, in case of CCPs in argon, which is perhaps the most studied gas by PIC/MCC simulations, electrons and (singly charged) argon ions are traced in the discharge gap. Other plasma species, such as metastables and fast neutrals created in ion-atom collisions are usually not taken into account in the models. 

Regarding the description of the secondary electron emission processes taking place at the boundary surfaces, further simplifications are also customary. In several studies secondary electron emission from the electrodes has simply been neglected. While CCPs can be sustained in their $\alpha$-mode of operation without secondary electrons \cite{Belenguer_1990,Schulze_Heat1,Schulze_Heat2,Schulze_Heat3,Turner_Heat,Mussenbrock_Heat,Schweigert_Heat,Ivanov_Heat}, this assumption is not always justified. In most studies that consider secondary electron emission (i) a constant value for the secondary electron yield, $\gamma$, is used, that is independent of the discharge conditions (e.g., the energy of impacting ions), (ii) only the ion-induced secondary electron emission is taken into account, neglecting the contributions of other plasma species, (iii) the effect of the surface conditions is not accounted for. In contrast to the $\alpha$-mode, beyond the mode transition to $\gamma$-mode, secondary electron emission plays an essential role in the ionization dynamics \cite{Belenguer_1990,Donko_2010_secondarye,Schulze_2011_secondarye,Turner_secondarye,Booth_secondarye,Boehm_secondarye,Liu_secondarye,brz_secondarye}. 

It is known that besides positive ions the fast neutrals, metastable atoms and VUV photons can as well contribute to secondary electron emission and that the importance of these species depends to a great extent on the discharge conditions (incident particle energies) and electrode surface properties (see  \cite{Phelps_1999}). The effect of these ``other'' species can implicitly be included in a discharge model via definition of an ``apparent'' or ``effective'' secondary electron emission coefficient (or secondary electron yield), as the ratio of the secondary electron flux to the ion flux at the electrode. The effective secondary electron emission yield, $\gamma^\ast$, has been obtained by Phelps and Petrovi\'c \cite{Phelps_1999} for the case of a homogeneous electric field (breakdown / Tonwsend discharge conditions) and by Donk\'o \cite{Donko_2000,Donko_2001} and Mari\'c {\it et al.} \cite{Maric} for cathode fall conditions in abnormal DC glow discharges in argon. 

The effects of including fast neutrals and realistic secondary electron yields in the calculations have already been addressed in several previous studies for specific geometries and discharge conditions, but are widely ignored in most current simulations of low pressure CCPs under conditions relevant for plasma processing applications. A series of simulation studies by Bogaerts {\it et al.} \cite{Bogaerts_1995b,Bogaerts_1996,Bogaerts_2002,Bogaerts_1999a,Bogaerts_1999b} on low-pressure DC and radio frequency (RF) analytical glow discharges in argon have demonstrated the importance of electrode surface conditions and fast neutrals contributing to the ionization in the gas phase and to the sputtering of the electrodes. Braginsky {\it et al.} \cite{Braginsky_2012} have found the contribution of fast atoms to the secondary electron emission to be comparable to the secondary electron emission due to ion impact in a low-frequency (1.76 MHz) CCP. The importance of taking account of heavy particle collisions in the calculations has been pointed out also in the case of oxygen and hydrogen RF plasmas \cite{A1,A2,A3,A4}. Bojarov {\it et al.} \cite{Bojarov} have recently studied the effects of energy dependent $\gamma$-coefficients and fast atoms in CCPs and the effects of different surface conditions at the powered and grounded electrodes. Secondary electrons also significantly affect the realization of the separate control of the ion flux and the mean ion energy at the electrodes in dual-frequency capacitive RF discharges such as found in \cite{Donko_2010_secondarye,Schulze_2011_secondarye}. An asymmetry effect induced by the different electron emission properties of the two electrodes of CCPs (having unequal $\gamma$-coefficients at both electrodes), reported by Lafleur {\it et al.} in \cite{Lafleur_2013_SEA}, was found to significantly influence the electrical generation of the DC self-bias and the independent tuning of ion properties in electrically asymmetric discharges \cite{Korolov_2013_SEA,UCZ_EAE_2009}.

These previous observations show that special attention must be paid to the set of plasma particles traced in PIC/MCC simulations of CCPs and to the precise description of the processes (taking place both in the discharge volume and at the boundary surfaces) that are implemented in the model, in order to achieve a realistic description of capacitive RF discharges.

Here, we perform a systematic investigation of the effects of fast neutrals and realistic energy-dependent secondary electron emission coefficients on the calculated discharge characteristics resulting from PIC/MCC simulations of CCPs under conditions relevant for plasma processing applications. We focus on single-frequency discharges driven at $f$=13.56 MHz and at three different pressures of 5 Pa, 20 Pa and 100 Pa to probe a non-local collisionless, an intermediate, and a collisional regime. At each of these pressures, simulations are carried out for a wide range of voltage amplitudes. The tracing of fast neutrals is switched on and off and different implementations of secondary electron emission from the electrodes due to heavy particle impact are included in the computations. Simulations with constant, as well as energy-dependent emission coefficients are performed and secondary electron emission is switched on and off. In this way gas phase and surface effects of heavy particles on the discharge characteristics are identified and separated. We find a moderate effect of the gas phase reactions, but a more significant effect of the secondary electron emission coefficient (even in the $\alpha$-mode of operation of the discharge at low pressures) on process relevant plasma parameters such as the plasma density and ion fluxes to the electrodes. 

In section 2, we describe the discharge conditions and specify different physical models that allow the identification of the above effects. The results are presented in section 3, which is split into 3 parts according to the 3 different pressures investigated (5 Pa, 20 Pa, and 100 Pa). Conclusions are drawn in section 4.

\section{Physical models and simulation method}

The calculations are based on our electrostatic 1d3v bounded plasma Particle-in-Cell code complemented with Monte Carlo treatment of collision processes (PIC/MCC) \cite{Donko_2011_PSST, Donko_2012_PPCF}, which is extended to handle additional processes to be discussed later on.

The discharges investigated are geometrically symmetric. The plane, parallel, and infinite electrodes, separated by a distance of 2.5 cm, are assumed to be made of the same material with identical surface conditions, hence characterized by the same electron emission and particle reflection properties. 
We cover neutral gas pressures of 5 Pa (low pressure), 20 Pa (intermediate pressure), and 100 Pa (high pressure). The neutral gas temperature is constant, taken to be 350 K. A voltage waveform of $V(t)=V_0 \cos(2\pi f t)$ with $f = 13.56$ MHz is applied to one electrode located at $x = 0$ cm, while the other electrode is grounded. 

At the electrodes, electrons are reflected with a probability of 0.2, independently of their energy and angle of incidence. This value is adopted from \cite{Kollath_erefl}. Secondary electron emission due to electron impact at the electrodes is neglected. These two assumptions are clearly  simplifications in our work, and follow the practice of most PIC/MCC simulations. In reality, the efficiencies of the reflection of  electrons and the creation of secondary electrons are known to depend on the incident electron energy and the angle of incidence (such as outlined by, e.g., Braginsky {\it et al.} \cite{Braginsky_2012}), as well as on the electrode materials \cite{Lin_2005,Scholtz_1996}. A systematic investigation of the effect of using realistic electron reflection coefficients will be addressed in a future paper.

\begin{table}[h]
\caption{Characteristics of the different models used in this work. $\gamma^\ast$ is the effective secondary electron yield calculated according to equation (\ref{gamma_en}). Ar$^{\rm f}$ denotes fast atoms.\\}
\label{table1}
\begin{tabular}{|l|l|l|l|}
\hline
 Model  & Secondary emission yield & PIC species & Collisions \\ \hline
A & $\gamma=0$ & \multirow{3}{*}{e$^-$, Ar$^+$} & 
\multirow{3}{*}{\begin{tabular}[c]{@{}l@{}}e$^-$+Ar, Ar$^+$+Ar\\ (elastic, excitation, 
ionization)\end{tabular} } \\ \cline{1-2}
B & $\gamma=0.1$ &  &  \\ \cline{1-2}
C & $\gamma=\gamma^\ast$ & & \\ \hline
D & $\gamma=0$ & \multirow{2}{*}{e$^-$, Ar$^+$, Ar$^{\rm f}$} & \multirow{2}{*}
{\begin{tabular}[c]{@{}l@{}}e$^-$+Ar, Ar$^+$+Ar, Ar$^{\rm f}$+Ar\\ 
(elastic, excitation, ionization)\end{tabular}} \\ \cline{1-2}
E & $\gamma=\gamma^\ast$ & &  \\ \hline
\end{tabular}
\end{table}

The different models used here are listed in Table 1. In the first set of the  models, A, B, and C, the ``active'' species of the PIC/MCC simulations are electrons and Ar$^+$ ions. The MC collision routine handles collisions of these species with the atoms of the background gas. In these models different approaches are used for the secondary electron emission:  model A neglects this process by setting $\gamma=0$,  model B uses $\gamma=0.1$, a value often adopted in discharge simulations, while  model C calculates the effective secondary electron yield based on the energies and corresponding yields of the individual positive ions impacting the electrodes. 

In the second set of the models, D and E, tracing of fast Ar atoms (Ar$^{\rm f}$) is also included. The fast neutrals are created mainly in the sheaths, as a result of elastic ion - thermal atom and subsequent fast atom - thermal atom collisions. Fast atoms are defined here as the ones having a kinetic energy above a threshold value, $\varepsilon_{\rm f}$. In most of our calculations this value is set to $\varepsilon_{\rm f} = $23 eV, except in the analysis of the velocity distribution functions of heavy particles, where the threshold is set at a lower value. The $\varepsilon_{\rm f} = $23 eV energy is near the threshold for impact excitation of the background gas by atoms having the same mass.  (Note that the energy available for an inelastic process (involving particles with equal masses) in the center-of-mass frame of reference is half of the projectile energy if the target particle is at rest. Thus, the threshold energy for the excitation of Ar atoms by fast neutrals is higher by a factor of two compared to the electron impact excitation threshold energy of 11.55 eV.) Fast atoms are traced in the gap until their energy drops below the threshold value, or until they arrive at the electrodes. 

For the conditions covered here ions and fast neutrals contribute dominantly to secondary electron emission \cite{Phelps_1999,Donko_2000}, thus we disregard the contributions of metastable atoms and UV/VUV photons. The effective secondary electron yield is calculated as:
\begin{equation}
\gamma^\ast= \frac{ \sum\limits_{k=1}^{N_{i}} \gamma_{i}(\epsilon_{k}) + \sum\limits_{k=1}^{N_{a}} \gamma_{a}(\epsilon_{k}) }{N_{i}},
\label{gamma_en}
\end{equation}
where $\epsilon_{k}$ is the energy of the ion or atom (upon arrival at the electrode) noted by $k$, $N_{i}$ and $N_{a}$ are the total number of ions and fast atoms reaching a given electrode during a RF period. 

The concept of using an effective secondary electron yield -- as already mentioned above -- follows Ref. \cite{Phelps_1999}. It is important to recognize that this coefficient corresponds to the number of electrons emitted per ion reaching the electrode, while other particles (in our case fast neutrals) also contribute to electron emission. This is why only $N_i$ appears in the denominator of eq. (\ref{gamma_en}). The effective secondary electron yield, $\gamma^\ast$, obtained via this definition, can differ significantly for various physical settings (DC cathode fall conditions \cite{Donko_2000,Donko_2001}, homogeneous electric fields conditions (e.g., Townsend discharges) \cite{Phelps_1999}, and RF discharges studied here), due to the specific particle dynamics and electric field distributions, even when the same gas - electrode material pair is considered.  

The energy-dependent secondary electron emission yields for ions and fast atoms, $\gamma_{i}$ and $\gamma_{a}$, respectively, used in this calculation are given as \cite{Phelps_1999,Phelps_Donko_1999}:
\begin{equation}
\gamma_{i}(\epsilon)= \frac{0.006\epsilon}{1+(\epsilon/10)} + \frac{1.05\times 10^{-4}(\epsilon-80)^{1.2}}{(1+\epsilon/8000)^{1.5}} ,
\label{gamma_i}
\end{equation}
\begin{equation}
\gamma_a(\epsilon)= \frac{1 \times 10^{-4}(\epsilon-90)^{1.2}}{1+(\epsilon/8000)^{1.5}} + \frac{7\times 10^{-5}(\epsilon-32)^{1.2}}{1+(\epsilon/2800)^{1.5}}.
\label{gamma_a}
\end{equation}
Note that a correction to ({\ref{gamma_i}), which appeared originally in \cite{Phelps_1999}, was given subsequently in \cite{Phelps_Donko_1999} and that we use coefficients that characterize surfaces typical for laboratory conditions (termed as ``dirty surfaces'' in \cite{Phelps_1999}). In the models where fast neutrals are not considered, the calculation of the apparent yield according to ({\ref{gamma_en}) uses only the first term of the numerator of the right-hand side of the equation.

\begin{figure}[h!]
\begin{center}
\includegraphics[width=0.5\textwidth]{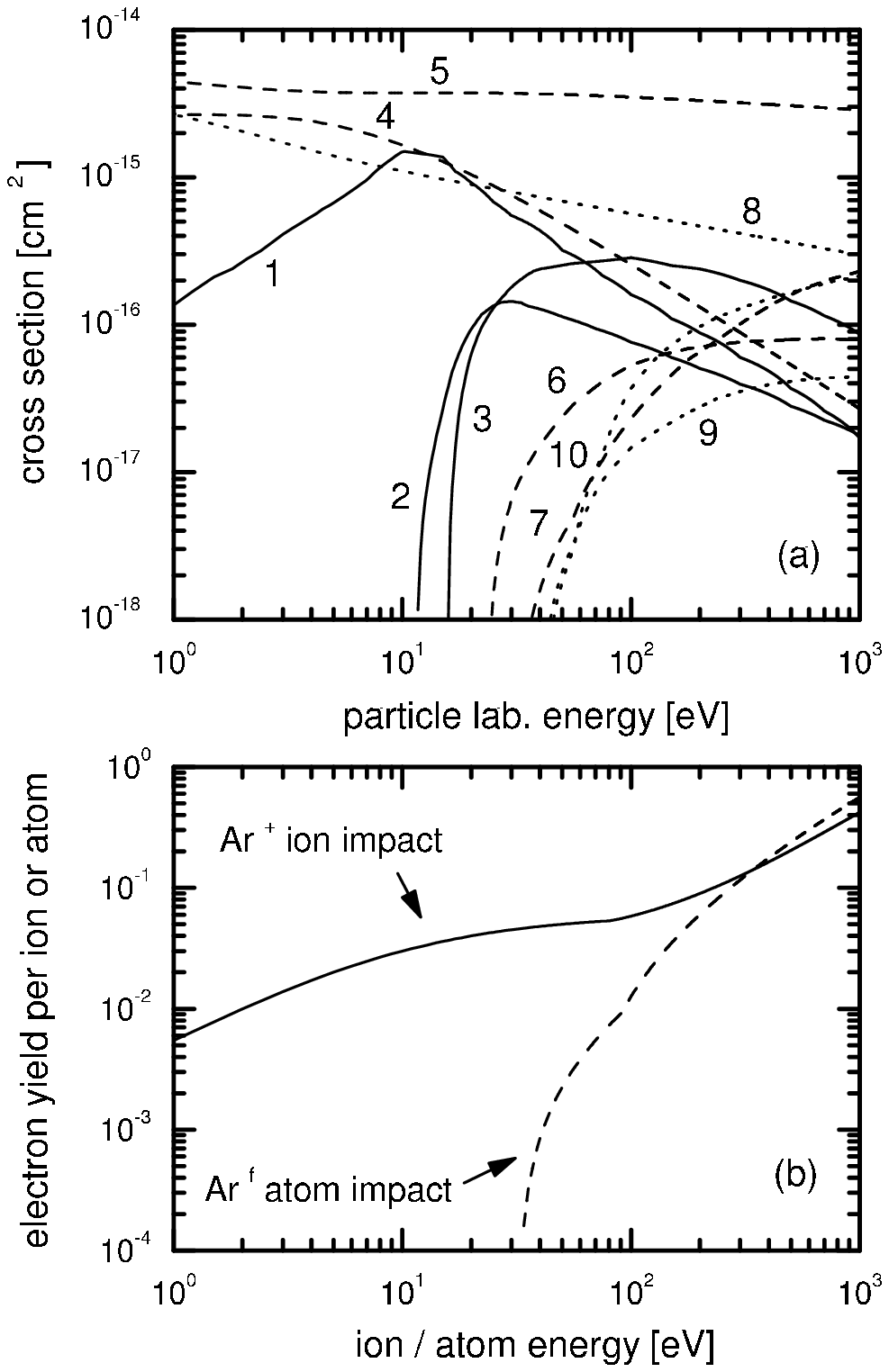}
\caption{(a) Cross sections of elementary processes used in the simulation \cite{Phelps_CS_jila,Phelps_CS_1991,Phelps_CS_1994}. The solid lines indicate electron collisions (1: elastic, 2: excitation, 3: ionization), the dashed lines indicate Ar$^+$ cross sections (4: isotropic part of elastic scattering, 5: backward elastic scattering, 6: excitation, 7: ionization), and the dotted lines indicate fast Ar atom cross sections (8: isotropic elastic scattering, 9: excitation, 10: ionization). (b) Energy dependence of secondary electron emission yields due to Ar$^+$ and fast Ar atom (Ar$^{\rm f}$) impact onto a copper electrode under typical laboratory conditions (termed as "dirty" surfaces in \cite{Phelps_1999}). }
\label{fig:Crosssections}
\end{center}
\end{figure}

The cross sections for electron-neutral and ion-neutral collision processes are taken from \cite{Phelps_CS_jila,Phelps_CS_1991,Phelps_CS_1994}, while for Ar$^+$+Ar and Ar$^{\rm f}$+Ar collisions (elastic scattering, ionization, and the dominant excitation processes) the source of cross section data is \cite{Phelps_CS_jila}. The set of cross sections is plotted in figure~\ref{fig:Crosssections}(a), while the secondary electron emission coefficients for ion and atom impact are plotted as a function of the incident particle energy in figure~\ref{fig:Crosssections}(b).

At each of the pressures covered, the effects of fast atoms and that of using energy-dependent secondary electron yields in the simulations are investigated. In order to clarify the effect of fast neutrals {\it in the gas phase}, simulation results obtained by (i) tracing only ions and (ii) tracing both ions and fast atoms are compared, while the secondary electron emission is neglected ($\gamma$ = 0) -  models A and D. To study the effect of considering energy-dependent secondary yields in the model, simulations are carried out and the results are compared for the following conditions: (i) only ions are traced and a constant secondary electron emission coefficient, $\gamma$ = 0.1, is used -  model B; (ii) only ions are traced and an energy-dependent secondary yield for ions is used -  model C; and (iii) both ions and fast neutrals are traced and energy-dependent secondary yields for these species are used -  model E. The last setting represents the most complete model that includes {\it both the gas-phase and surface effects} of the heavy particles, Ar$^+$ ions and fast neutral atoms, in a realistic way, by calculating $\gamma^\ast$ based on elementary data on energy-dependent secondary electron yields of these two species. 

We note that including fast neutrals in the computations results in a marginal increase of the computation time, by only 3 \% - 5 \% for our conditions, as long as the threshold energy $\varepsilon_{\rm f}$ has the high value of 23 eV, as fast atoms can be traced with the same simulation time step as the ions, for which a ``subcycling'' approach is used (i.e., ions and fast atoms are moved in every 20th simulation step). When a lower threshold energy is set (as it would be required, e.g., for gas heating calculations), the number of neutrals to be traced and the computational time increase significantly.

\section{Results}

\subsection{Low pressure (5 Pa)}

\begin{figure}[ht!]
\begin{center}
\includegraphics[width=1.0\textwidth]{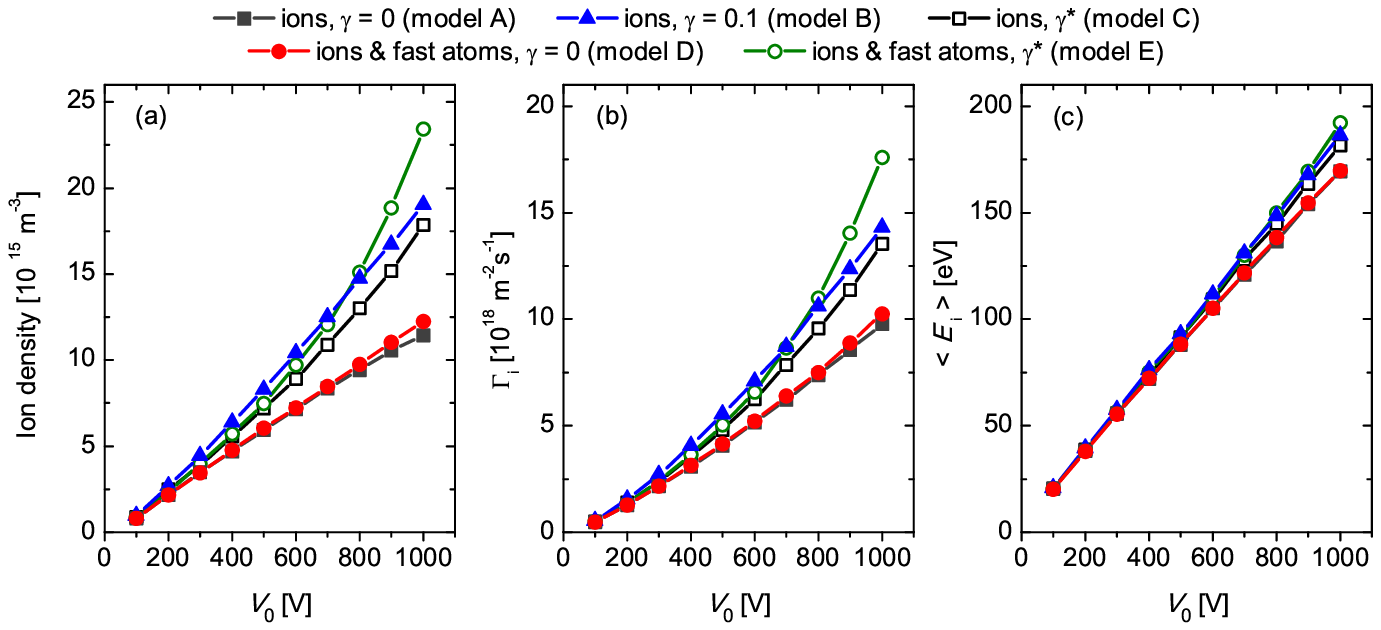}
\caption{Ion density in the center of the discharge (a), ion flux, $\Gamma_{\rm{i}}$ (b), and mean ion energy, $\langle E_{\rm{i}} \rangle$ (c), at the electrodes as a function of the driving voltage amplitude, obtained from PIC/MCC simulations based on  models A -- E. Discharge conditions: 5 Pa, 13.56 MHz, 2.5 cm electrode gap.}
\label{fig:iondensity_5Pa}
\end{center}
\end{figure}

In Figure \ref{fig:iondensity_5Pa}, simulation results for the ion density in the center of the discharge (a), the flux (b) and the mean energy (c) of ions at the electrodes are plotted as a function of the driving voltage amplitude at 5 Pa, based on  models A -- E. At low voltage amplitudes ($V_0$) all the models predict very similar values for the above characteristics. Differences up to a factor of two are found between the results of the calculations based on the different models at the highest voltage ($V_0$ = 1000 V) for the ion density and ion flux, whereas the mean ion energy at the electrodes proves to be rather insensitive of the modeling assumptions. 

The lowest densities and fluxes at all voltage values are computed when the secondary electron yield is set to zero. The slightly higher density and flux obtained with  model D, as compared to  model A, results from a gas-phase effect: ionization by fast heavy particles (ions and neutrals). Including a secondary electron yield $\gamma >0$ increases the ion density and flux:  models B and C, assuming a constant ($\gamma =0.1$) secondary electron yield and a calculated electron yield $\gamma^\ast$ by considering ions only, respectively, result in about 60\% higher values than  models A and D (which both assume $\gamma =0$). The contribution of fast neutrals to secondary electron emission further increases the ion density and flux as indicated by the results of  model E. This behavior can be explained by the increase of the secondary electron yields of ions and fast neutrals with their energy, on which the driving voltage has a great influence at the low collisionality of the sheaths at 5 Pa gas pressure. 

\begin{figure}[ht!]
\begin{center}
\includegraphics[width=0.75\textwidth]{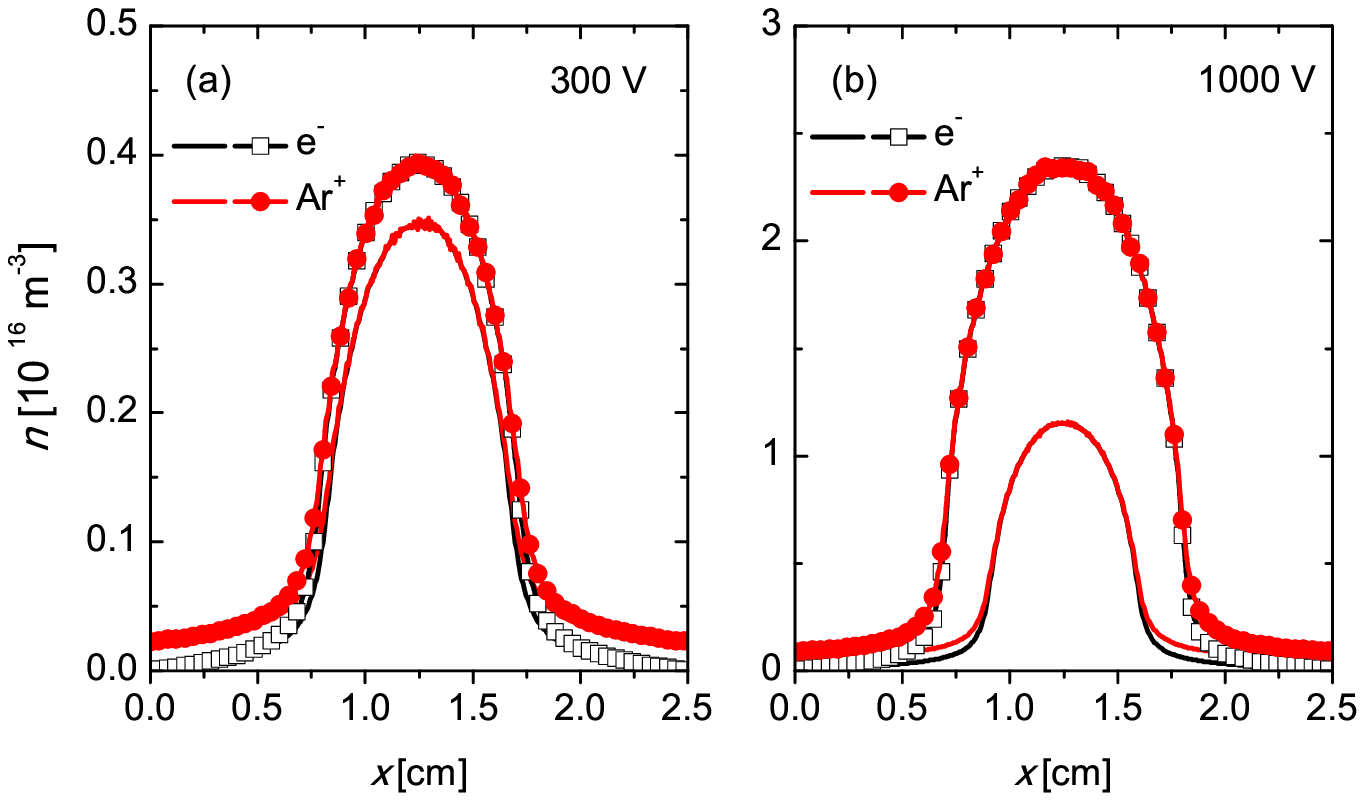}
\caption{Time-averaged charged particle density distributions for different voltage amplitudes, (a) $V_0$=300 V and (b) $V_0$=1000 V, obtained from PIC/MCC simulations based on  model A (lines) and  model E (lines with symbols). Discharge conditions: 5 Pa, 13.56 MHz, 2.5 cm electrode gap.}
\label{fig:dens_5Pa}
\end{center}
\end{figure}

It is important to note that an accurate description of the secondary electron emission processes proves to be quite important under these low pressure conditions, despite the fact that such discharges are usually quoted to operate in the $\alpha$-mode, where sheath-expansion heating dominates (see below). In particular, the effects of secondary electrons are found to be important at the high driving voltages, which is the domain of operation of plasma processing applications that require a high plasma density.   

Figure \ref{fig:dens_5Pa} displays the temporally averaged charged particle density profiles in the discharge, as obtained from computations based on  models A and E. At the lower voltage of $V_0$ = 300 V the densities grow by about 10\% when the most complete physical model is used (model E), compared to the results obtained on the basis of the simplest model that considers only electrons and ions, and neglects secondary electron emission from the electrodes (model A). At the higher voltage amplitude, $V_0$ = 1000 V, however, these differences become significant and amount a factor of two, indicating pronounced effects of gas-phase processes of fast heavy particles (ionization) and of secondary electron emission from the electrodes. Besides the differences of the peak densities observed here, figure \ref{fig:dens_5Pa}(b) also reveals a remarkable difference in the sheath length obtained from the different models.  

\begin{figure}[ht!]
\begin{center}
\includegraphics[width=0.95\textwidth]{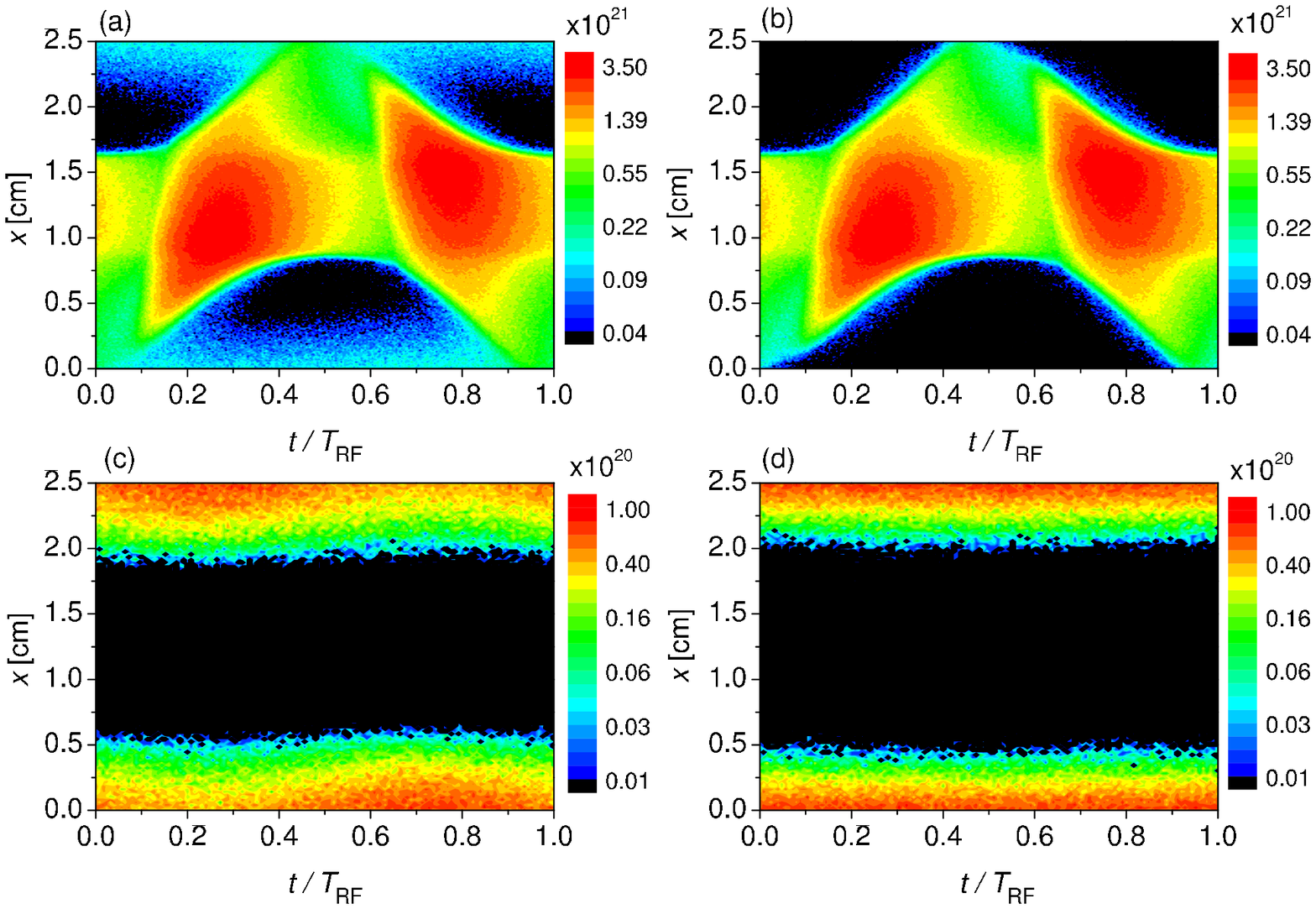}
\caption{Spatio-temporal plot of the total ionization rate (a) and the contribution of electrons (b), ions (c) and fast neutrals (d) to the ionization, obtained from PIC/MCC simulations based on model D, i.e. by tracing both ions and fast neutrals in the model. Discharge conditions: 5 Pa, 13.56 MHz, 2.5 cm electrode gap, $V_0$=1000 V, $\gamma$ = 0. The color scales are logarithmic, cover two orders of magnitude, and are given in units of m$^{-3}$s$^{-1}$.}
\label{fig:ionization_rates_5Pa}
\end{center}
\end{figure}

Figure \ref{fig:ionization_rates_5Pa} shows spatio-temporal plots of the total ionization rate (a) and the individual contributions of electrons (b), ions (c), and neutrals (d) for a voltage amplitude of 1000~V, as resulting from  model D (including fast neutrals, but $\gamma = 0$), as an example. The powered electrode is at $x$ = 0.0 cm and the grounded electrode is at $x$ = 2.5 cm. Under these conditions, the discharge operates in the $\alpha$-mode, i.e. the ionization is dominated by electrons heated by sheath expansion. The ionization rate associated with the fast neutrals and ions is in the same order of magnitude and is appreciable only near the electrodes. The comparable ionization rate by the two species is a result of similar fluxes and ionization cross sections (figure \ref{fig:Crosssections}). The localization is explained by the fact that ions are accelerated towards the electrodes by the sheath electric field resulting in higher probabilities for an ionizing collision with the neutral background gas close to the electrodes. Fast neutrals are mainly produced by charge exchange collisions of ions with thermal neutrals. Due to the acceleration of positive ions towards the electrode this results in faster neutrals in close vicinity to the electrode and, therefore, more ionization by neutrals at the electrode compared to further away from boundary surfaces. Fast neutrals also have an indirect, but important effect on the electron impact ionization rate: they cause ionization inside the sheaths, i.e., they generate electrons inside the sheaths. Similarly to secondary electrons generated at boundary surfaces, these electrons are accelerated towards the plasma bulk by the sheath electric field and can be multiplied by collisions. In this way ionization by fast neutrals increases the electron impact ionization rate (we note that a similar effect takes place in the sheath of cold-cathode DC discharges \cite{Donko_2001}). At the low pressures of 5 Pa the collisional multiplication of electrons inside the sheaths is inefficient and the effect of the ionization by fast neutrals on the ionization rate of electrons is relatively weak, but present. This effect is more important at higher pressures. We also note that while the ionization rate corresponding to ions, figure \ref{fig:ionization_rates_5Pa}(c), shows a modulation in time, the fast neutral induced ionization rate (shown in figure \ref{fig:ionization_rates_5Pa}(d)) exhibits no time dependence. The temporal dependence of the ionization attributed to ions is related to the modulation of the ion flux by the time-varying electric field in the regions near the electrodes.

\begin{figure}[ht!]
\begin{center}
\includegraphics[width=0.5\textwidth]{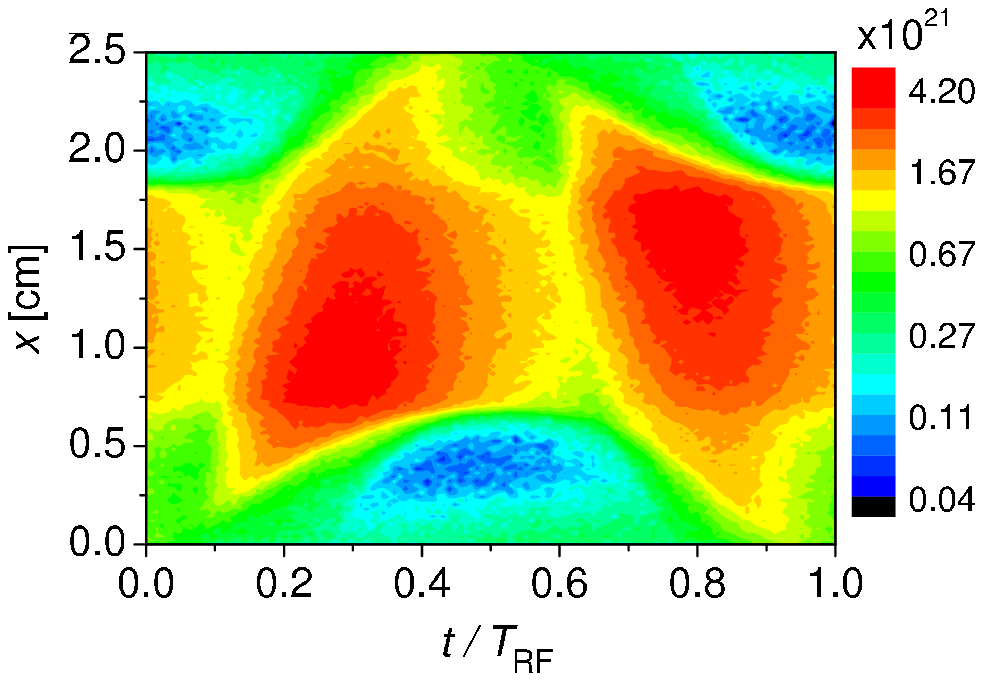}
\caption{Spatio-temporal plot of the total ionization rate as obtained from  model E for $p$ = 5 Pa and $V_0$ = 1000 V. The color scale is logarithmic, covers two orders of magnitude, and is given in units of m$^{-3}$s$^{-1}$.}
\label{fig:totion_5Pa}
\end{center}
\end{figure}

The total ionization rate for $V_0$ = 1000 V, as resulting from our most accurate model (model E, which includes the calculation of the effective secondary electron yield, $\gamma^\ast$) is displayed in figure~\ref{fig:totion_5Pa} for $p$ = 5 Pa and $V_0$ = 1000 V. The plot reveals the dominance of the $\alpha$-mechanism in the ionization, however, the main maxima of the ionization rate extend far beyond the expansion of the sheaths (c.f. also with figure \ref{fig:ionization_rates_5Pa}(a)) and indicate significant ionization in the bulk plasma at phases of expanded  sheaths. This contribution, as well as ionization within the sheaths near the sheath/bulk boundaries, are maintained by electrons  (emitted from, or created near the electrodes) that are accelerated in the sheath electric field. Additionally, one can also identify ionization near the electrodes, which is caused by collisions of fast heavy particles (as seen in panels (c) and (d) of figure \ref{fig:ionization_rates_5Pa}).

\begin{figure}[h!]
\begin{center}
\includegraphics[width=0.45\textwidth]{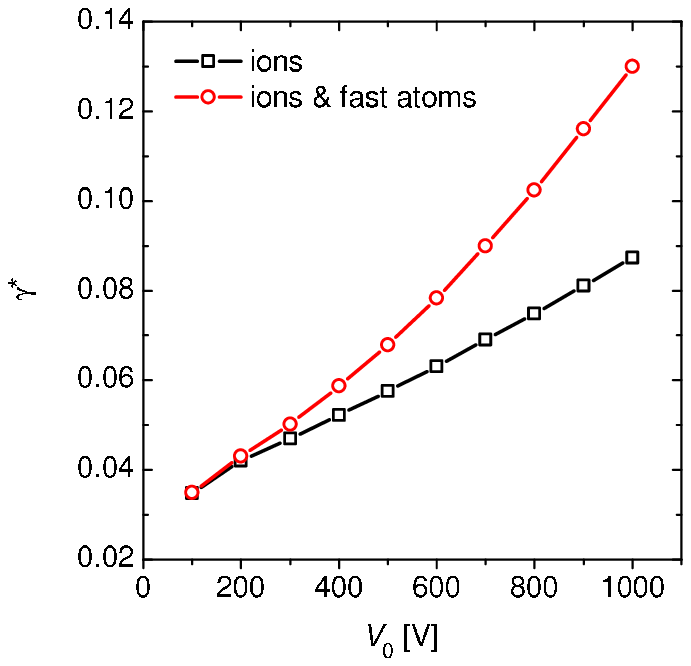}
\caption{Effective secondary electron emission coefficient as a function of the driving voltage amplitude obtained from PIC/MCC simulations using energy-dependent secondary electron emission coefficients and tracing only ions (line with open squares - model C) and both ions and fast neutrals (line with open circles - model E) in the simulations. Discharge conditions: 5 Pa, 13.56 MHz, 2.5 cm electrode gap. }
\label{fig:gamma_5Pa}
\end{center}
\end{figure}

In figure~\ref{fig:gamma_5Pa} we compare the effective secondary electron yield values obtained in simulations based on  models C (tracing only ions) and E (tracing ions and fast neutrals). The data are presented as a function of the driving voltage amplitude. At low voltages the contribution of the fast neutrals is negligible, since at such conditions the energy of fast neutrals at the electrodes is low and no secondary electrons are generated upon their impact (see figure \ref{fig:Crosssections}(b)). At higher voltage amplitudes both the ion and neutral induced electron yields rise and, therefore, $\gamma^\ast$ increases as a function of $V_0$. Meanwhile, the contribution of fast neutrals becomes clearly remarkable. At the highest voltage amplitude fast atoms in  model E increase $\gamma^\ast$ by about 50\%, compared to the results of  model C.
The significant change of the effective secondary electron yield $\gamma^\ast$ with discharge conditions (change with $V_0$ by about a factor of 2 for  model C and by almost a factor of 4 for  model E) directs the attention to the problem of using a constant yield for a wide range of conditions, as often used in simulation studies. It should be noted, however, that ``elementary'' energy-dependent secondary electron yield data, such as shown in figure \ref{fig:Crosssections}(b), are hardly available for different gas / electrode material combinations.

\subsection{Intermediate pressure (20 Pa)}

\begin{figure}[ht!]
\begin{center}
\includegraphics[width=1.0\textwidth]{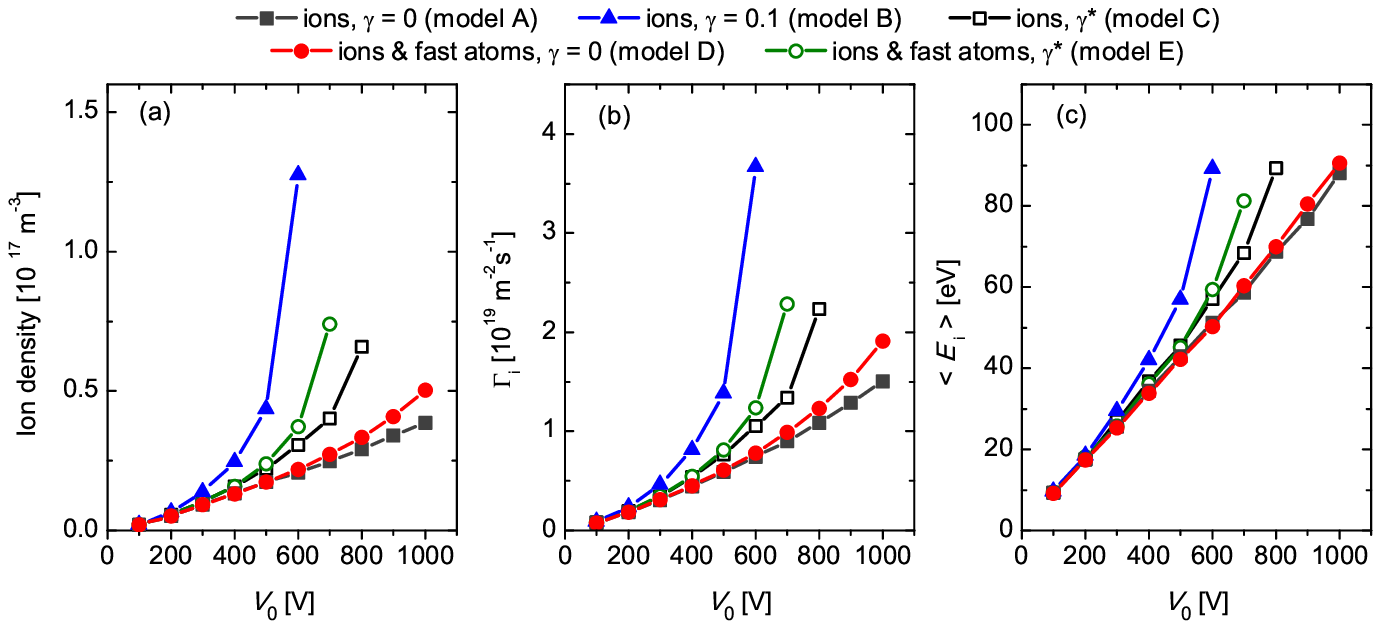}
\caption{Ion density in the center of the discharge (a), ion flux, $\Gamma_{\rm{i}}$ (b), and mean ion energy, $\langle E_{\rm{i}} \rangle$ (c), at the electrodes as a function of the driving voltage amplitude, obtained from PIC/MCC simulations with  models A -- E. Discharge conditions: 20 Pa, 13.56 MHz, 2.5 cm electrode gap.}
\label{fig:iondensity_20Pa}
\end{center}
\end{figure}

In figure~\ref{fig:iondensity_20Pa}, the ion density at the center of the discharge, as well as the flux and mean energy of ions reaching the electrodes are shown as a function of the driving voltage amplitude for $p$ = 20 Pa. The effect of the modeling assumptions on the discharge characteristics is more pronounced here compared to the low pressure, 5 Pa case (cf. figure \ref{fig:iondensity_5Pa}). The comparison of the results of models A and D, both of which neglect secondary electron emission, shows a moderate effect of heavy particle processes in the gas phase -- the increase of the plasma density and ion flux (due to ionization by heavy particles) is in the order of 20-30 \% at the highest driving voltage amplitude. The effect of fast atoms on the calculated plasma density can also be observed in figure~\ref{fig:densities_20Pa}, where the time-averaged charged particle densities are presented for 800~V and 1000~V. Figure~\ref{fig:densities_20Pa} reveals the impact of fast neutrals on the length of the sheath as well: tracing fast neutrals in the model results in a decrease of the sheath widths, which can be explained based on the ionization dynamics. 

Figure \ref{fig:ionization_rates_20Pa} shows spatio-temporal plots of the total ionization rate (a) and the individual contributions of electrons (b), ions (c), and neutrals (d) for a voltage amplitude of 1000~V, as resulting from model D (including fast neutrals, but $\gamma = 0$). Fast neutrals cause ionization close to the electrodes. Moreover, at 20 Pa the effect of ionization by fast neutrals on the electron impact ionization rate is much stronger compared to the low pressure scenario discussed in the previous section for two reasons: (i) more electrons are generated inside the sheath by ionization induced by fast neutrals and (ii) these electrons are effectively multiplied in the sheath. These electrons are accelerated to high energies in the sheath and generate ionization in the bulk, which finally leads to the increase of the plasma density and decrease of the length of the sheath. This effect is stronger at higher driving voltage amplitudes due to more ionization by fast neutrals inside the sheaths and a more effective acceleration and multiplication of electrons generated inside the sheaths by heavy particle ionization.

\begin{figure}[ht!]
\begin{center}
\includegraphics[width=0.75\textwidth]{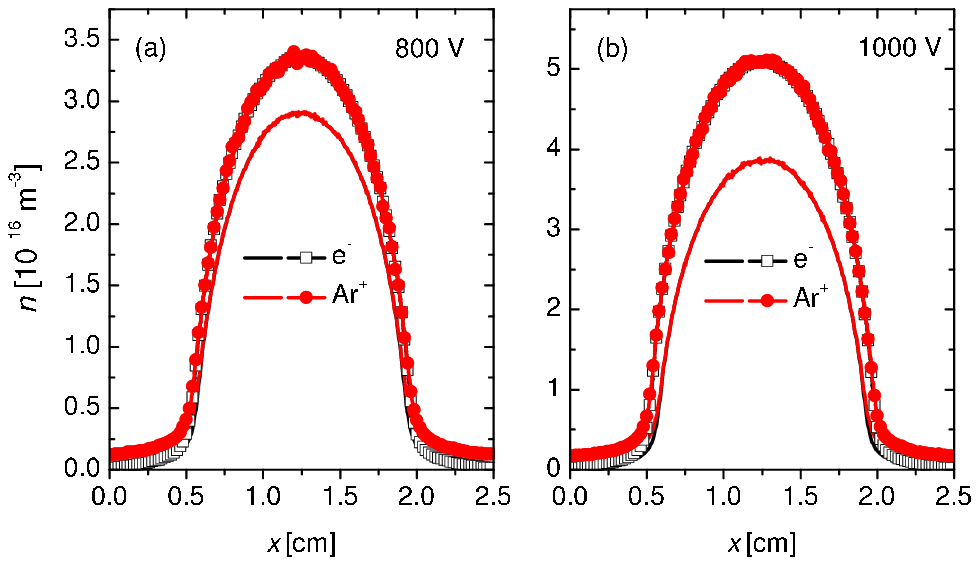}
\caption{Time-averaged charged particle density distributions for different voltage amplitudes, (a) $V_0$=800 V and (b) $V_0$=1000 V, obtained from PIC/MCC simulations by tracing only ions (lines - model A) and both ions and fast neutrals (lines with symbols - model D). Discharge conditions: 20 Pa, 13.56 MHz, 2.5 cm electrode gap, $\gamma$ = 0.}
\label{fig:densities_20Pa}
\end{center}
\end{figure}

\begin{figure}[ht!]
\begin{center}
\includegraphics[width=0.95\textwidth]{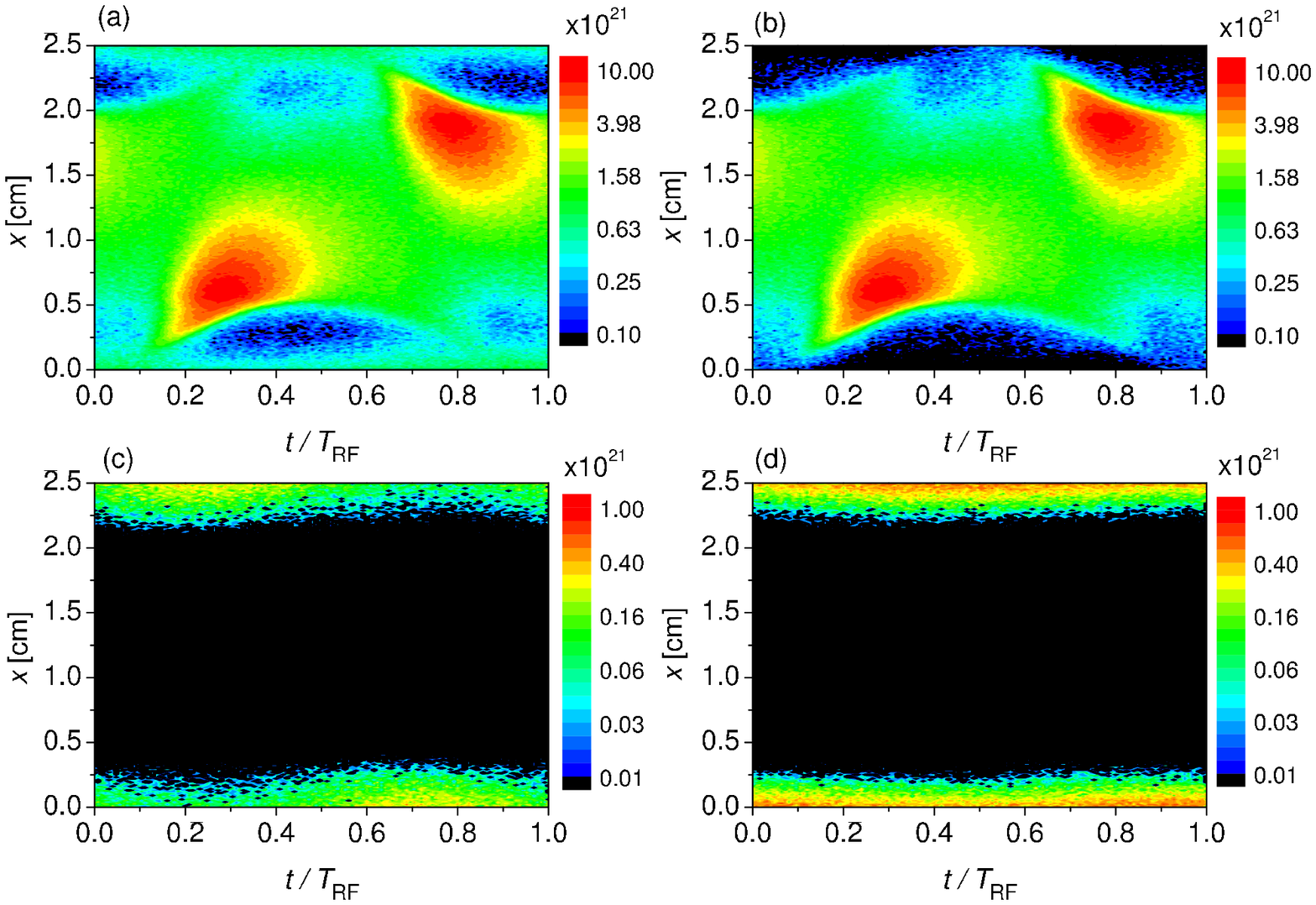}
\caption{Spatio-temporal plot of the total ionization rate (a) and the contribution of electrons (b), ions (c) and fast neutrals (d) to the ionization, obtained from PIC/MCC simulations based on model D, i.e., by tracing both ions and fast neutrals in the model. Discharge conditions: 5 Pa, 13.56 MHz, 2.5 cm electrode gap, $V_0$=1000 V, $\gamma$ = 0. The color scales are logarithmic, cover two orders of magnitude, and are given in units of m$^{-3}$s$^{-1}$.}
\label{fig:ionization_rates_20Pa}
\end{center}
\end{figure}

Figure~\ref{fig:iondensity_20Pa} also reveals that, compared to the cases (models A and D) with $\gamma=0$, the ion density, the ion flux and mean ion energy gradually increase (at any fixed voltage amplitude) by the inclusion of an energy-dependent secondary electron yield of the ions (model C), by considering additionally secondary electrons liberated by fast neutrals (model E), and by assuming a constant secondary yield of $\gamma=0.1$ (model B). This sequence suggests (as will be shown later) that the effective secondary electron yield $\gamma^\ast$ remains always below 0.1 for the conditions covered here. 

While the gas phase processes of fast atoms have a limited impact on the ion density (figure~\ref{fig:iondensity_20Pa}(a)), the secondary electrons emitted by fast atoms result in a significant increase of the ion density, at voltage amplitudes exceeding $\approx$ 500 V. The higher ion fluxes at these conditions (figure~\ref{fig:iondensity_20Pa}(b)) are the consequence of the higher plasma density.
The mean energy of ions at the electrodes, displayed in figure~\ref{fig:iondensity_20Pa}(c), is also affected by fast atoms. The increase of the mean energy is due to the effect of fast atoms on the length of the sheath (figure~\ref{fig:densities_20Pa}): when fast atoms are traced, higher plasma densities and shorter sheath lengths are obtained. The decrease of the sheath length leads to less collisions involving ions in the sheath. Therefore, ions reach the electrodes at higher energies.

\begin{figure}[ht!]
\begin{center}
\includegraphics[width=0.45\textwidth]{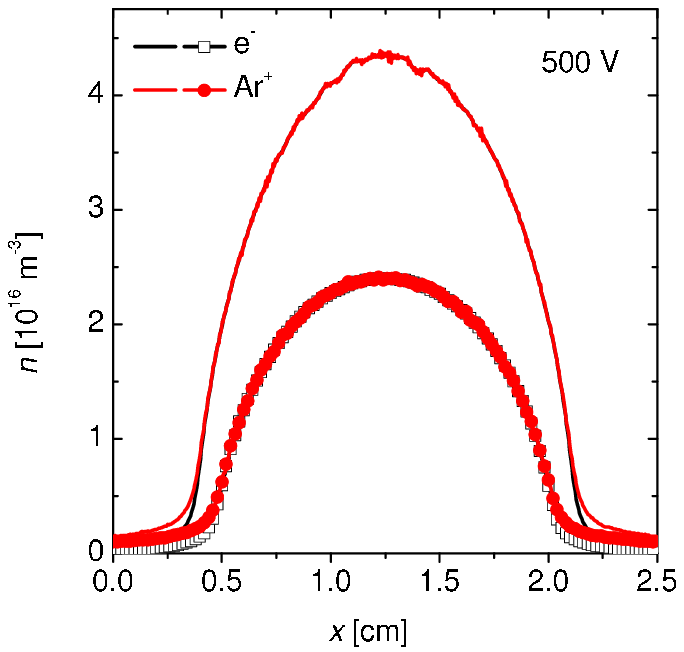}
\caption{Time-averaged charged particle density distributions obtained from PIC/MCC simulations using an effective secondary electron emission coefficient, $\gamma^\ast$, and tracing both ions and fast neutrals in the simulation (lines with symbols - model E), and using a constant secondary electron emission coefficient, $\gamma$ = 0.1, and tracing only ions in the simulation (lines - model B). Discharge conditions: 20 Pa, 13.56 MHz, 500 V voltage amplitude, 2.5 cm electrode gap.}
\label{fig:iondensity_20Pa_gen}
\end{center}
\end{figure}

A remarkable effect seen in the simulations at high voltage amplitudes is the divergence of the discharge characteristics for certain conditions. These conditions exclude models A and D -- in these settings the simulations converge for the whole domain of driving voltage amplitudes. In the other models, however, where the apparent secondary yield becomes high at high voltages (models C and E, see later) or is fixed at a high value (model B), convergence is found only for driving voltage amplitudes below $\sim$ 500 V. We note that in experiments, the driving voltage amplitudes are typically limited to about 500 V under these conditions as increasing the driving power results in an increase of the current, but the voltage increase is limited \cite{Godyak1,Godyak2}. This again shows that including fast neutrals in simulations of CCPs under conditions relevant for plasma processing applications is important.

A comparison of the time-averaged charged particle densities is shown in figure \ref{fig:iondensity_20Pa_gen}, for an effective secondary electron emission coefficient, $\gamma^\ast$, and tracing both ions and fast neutrals in the simulation (model E), and using a constant secondary electron emission coefficient, $\gamma$ = 0.1, and tracing only ions in the simulation (model B). In agreement with the results presented in figure \ref{fig:iondensity_20Pa}, it is observed that the densities of electrons and ions is increased and the width of the sheaths is decreased by changing from model E to model B.

This behavior can be understood based on the effective electron emission yield (calculated with models C and E), which is displayed in figure~\ref{fig:gamma_20Pa} as a function of the driving voltage amplitude. The calculated electron emission yield, $\gamma^\ast$, exhibits a strong dependence on the voltage amplitude, but always remains below the value of $\gamma=0.1$ (that is often assumed in simulations). At voltages $V_0 \leq$ 350 V the two models lead to nearly equal values of $\gamma^\ast$; both curves increase as a function of $V_0$ due to the increase of the energy-dependent secondary electron yield of ions. At higher voltages, when electron emission due to fast neutrals becomes important, the $\gamma^\ast$ obtained from model E increases more rapidly than that obtained on the basis of model C. The increase of $\gamma^\ast$ causes the divergence of the simulations at high values of $V_0$. Although the effective secondary electron yields here are lower than those obtained at 5 Pa, at 20 Pa secondary electrons contribute more significantly to the overall ionization due to their more efficient multiplication inside the sheaths. This is illustrated by figure \ref{fig:totion_20Pa} that shows a spatio-temporal plot of the total ionization rate obtained from  model E at 20 Pa and a driving voltage amplitude of 700 V. In contrast to figure \ref{fig:totion_5Pa} (5 Pa case), this plot shows much stronger ionization at the times of maximum sheath extension relative to the ionization during the phase of sheath expansion.

\begin{figure}[ht!]
\begin{center}
\includegraphics[width=0.45\textwidth]{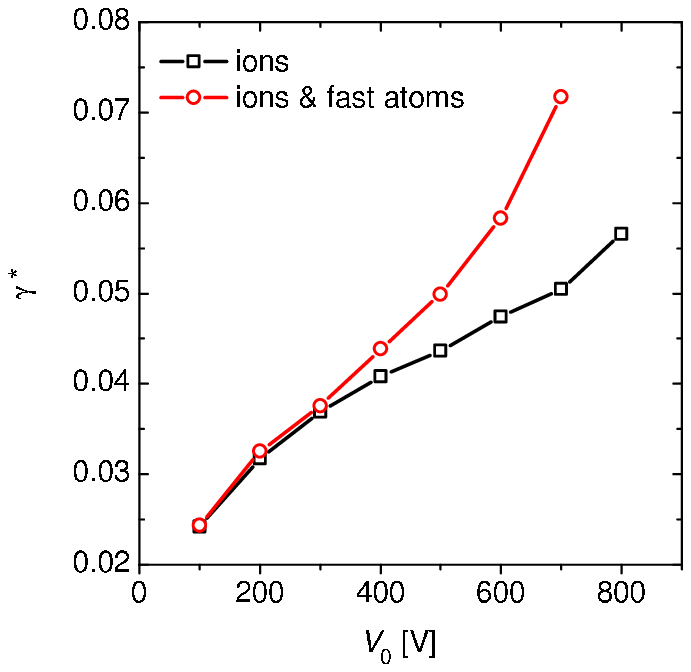}
\caption{Effective secondary electron emission coefficient, $\gamma^{*}$, as a function of the driving voltage amplitude obtained from PIC/MCC simulations using energy-dependent secondary electron emission coefficients and tracing only ions (line with open squares - model C) and both ions and fast neutrals (line with open circles - model E) in the simulations. Discharge conditions: 20 Pa, 13.56 MHz, 2.5 cm electrode gap. }
\label{fig:gamma_20Pa}
\end{center}
\end{figure}

\begin{figure}[ht!]
\begin{center}
\includegraphics[width=0.45\textwidth]{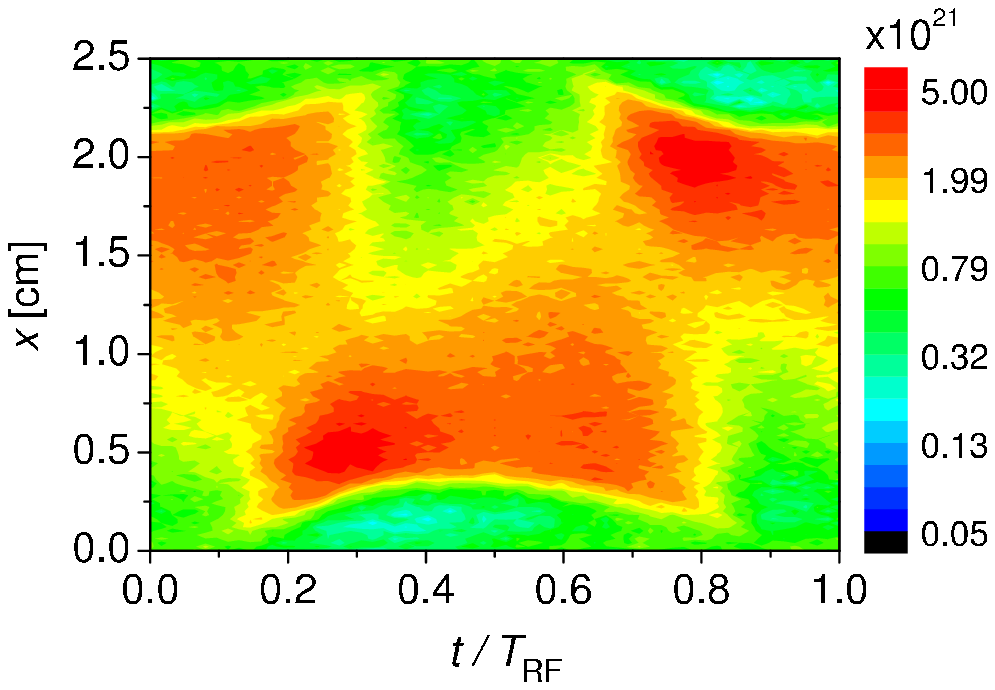}
\caption{Spatio-temporal plot of the total ionization rate as obtained from  model E for $p$ = 20 Pa and $V_0$ = 700 V. The color scale is logarithmic, covers two orders of magnitude, and is given in units of m$^{-3}$s$^{-1}$.}
\label{fig:totion_20Pa}
\end{center}
\end{figure}

For a more complete characterization of the motion of heavy particles we also analyze the time-averaged velocity distribution functions (VDF) of ions and fast neutrals. Due to the symmetry of the system the ``general'' VDF, $f({\bf r},{\bf v})$ reduces to $f(x,v_x,v_r)$ with a velocity component, $v_x$, along the discharge axis and a lateral velocity component, $v_r$. Figure~\ref{fig:vdf_20pa} shows $f(x,v_x,v_r)$ integrated over the 0.4 cm $<x<$ 0.6 cm spatial domain, situated inside the sheath region of the powered electrode (located at $x=0$). In these simulations the threshold energy for fast atoms was set to $\varepsilon_{\rm f}$ = 0.5 eV. The VDF of Ar$^+$ ions, shown in panel (a) indicates that the motion of ions is highly directional towards the powered electrode, as directed by the strong sheath electric field. The fast neutrals have a strongly anisotropic distribution, as well, at high velocities (energies) as these atoms originate from collisions with the highly directed ions. Following a sequence of atom-atom collisions, in which their energy decreases, a more isotropic distribution develops at lower energies. The white domain (circle) at low velocities corresponds to the threshold energy defined above. (Within this domain no information is available about the VDF.)

\begin{figure}[ht!]
\begin{center}
\includegraphics[width=0.95\textwidth]{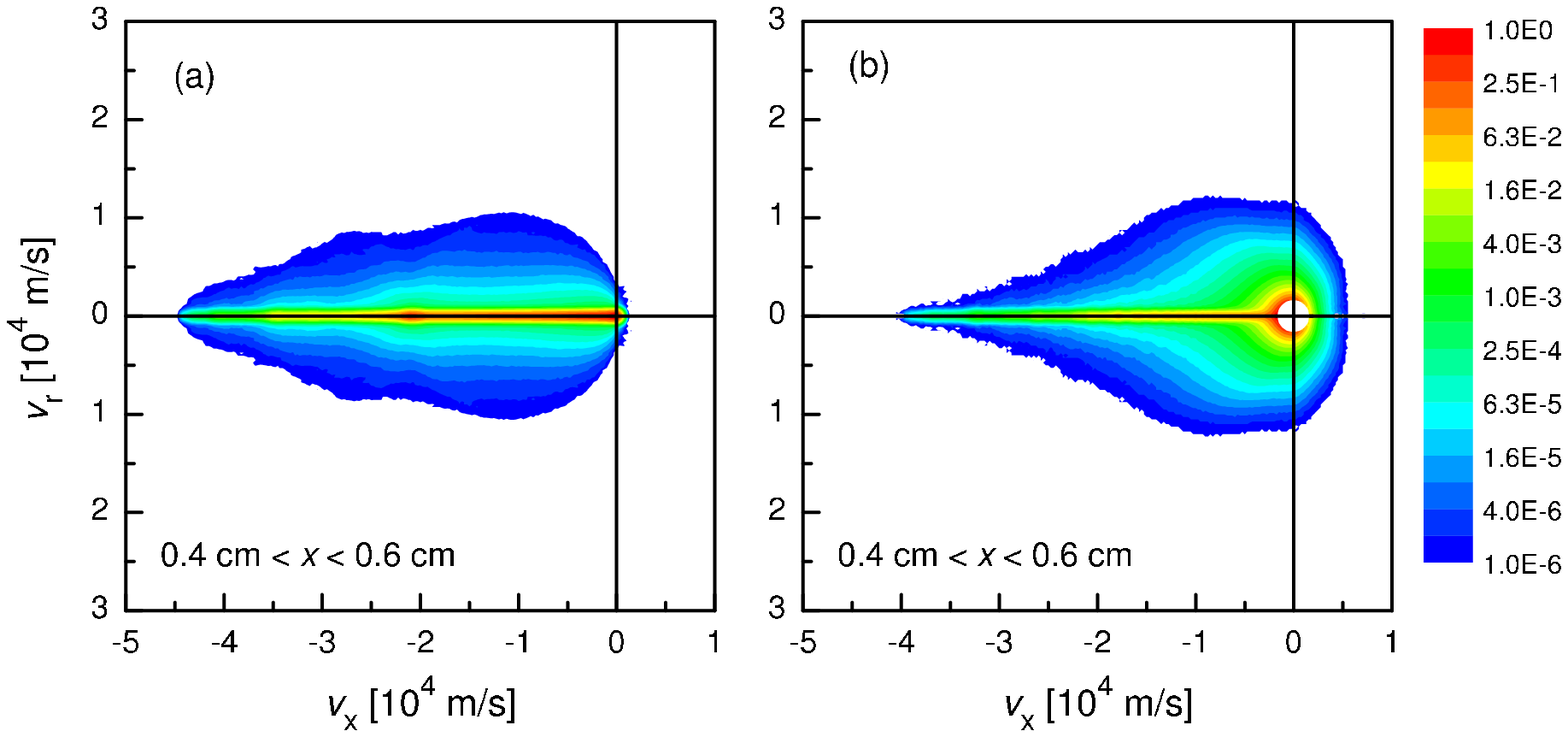}
\caption{$f(v_x,v_r)$ velocity distribution function of Ar$^+$ ions (a) and fast neutrals (b) in the sheath region of the powered electrode (situated at $x$ = 0), within the domain 0.2 cm $ < x < $ 0.4 cm. Here the limit energy for fast atoms was chosen to be 0.5 eV, the white region around $v=0$ in plot (b) corresponds to this limit. The color scale is given in arbitrary units, is logarithmic and covers 6 orders of magnitude.}
\label{fig:vdf_20pa}
\end{center}
\end{figure}

\subsection{High pressure (100 Pa)}

Figure~\ref{fig:iondensity_100Pa} shows the central ion density, as well as the flux and mean energy of ions reaching the electrodes, as a function of the driving voltage amplitude at 100 Pa for models A -- E. In this way a collisional regime is investigated, while collisionless and intermediate regimes were studied in the previous sections. Nevertheless, all trends are qualitatively similar to the 20 Pa case studied in the previous section: while a general increase in the ion density, flux, and energy is observed for increasing discharge voltage amplitudes, the differences between the different models becomes more severe at high voltages. Note that the highest driving voltage amplitudes, for which the simulation converges, are significantly lower in all models compared to the lower pressure scenarios. This is in agreement with experiments, where the current increases with increasing driving power, while the voltage remains low \cite{Godyak1,Godyak2}. By comparing  models A and D, we find that including fast neutrals has a minor effect on the ion properties. At this high pressure, the gas phase effects of tracing fast neutrals in the simulation are greatly reduced due to the highly collisional sheaths, so that the heavy particle energies inside the sheaths are small. Therefore, the additional ionization by fast neutrals and the subsequent enhancement of the electron impact ionization rate are low. 

\begin{figure}[h!]
\begin{center}
\includegraphics[width=1.0\textwidth]{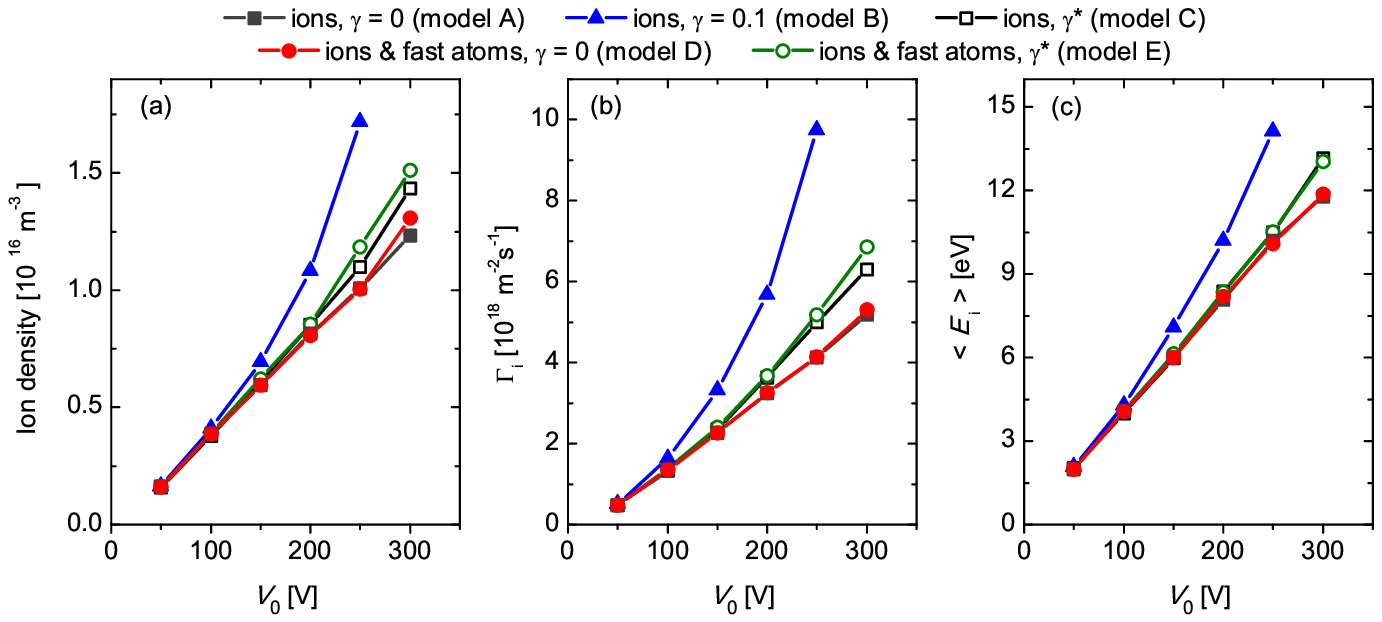}
\caption{Ion density in the center of the discharge (a), ion flux, $\Gamma_{\rm{i}}$ (b), and mean ion energy, $\langle E_{\rm{i}} \rangle$ (c), at the electrodes as a function of the driving voltage amplitude, obtained from PIC/MCC simulations, based on models A -- E. Discharge conditions: 100 Pa, 13.56 MHz, 2.5 cm electrode gap.}
\label{fig:iondensity_100Pa}
\end{center}
\end{figure}

Figure \ref{fig:densities_100Pa} shows the time averaged density profiles for electrons and ions resulting from models A and D (with and without fast neutrals) at 300 V. The effect of fast neutrals is strongest at the highest voltage amplitude of stable discharge operation in the high pressure regime, yet the changes in the charged particle densities and sheath widths stay well below 10 \%.

\begin{figure}[h!]
\begin{center}
\includegraphics[width=0.45\textwidth]{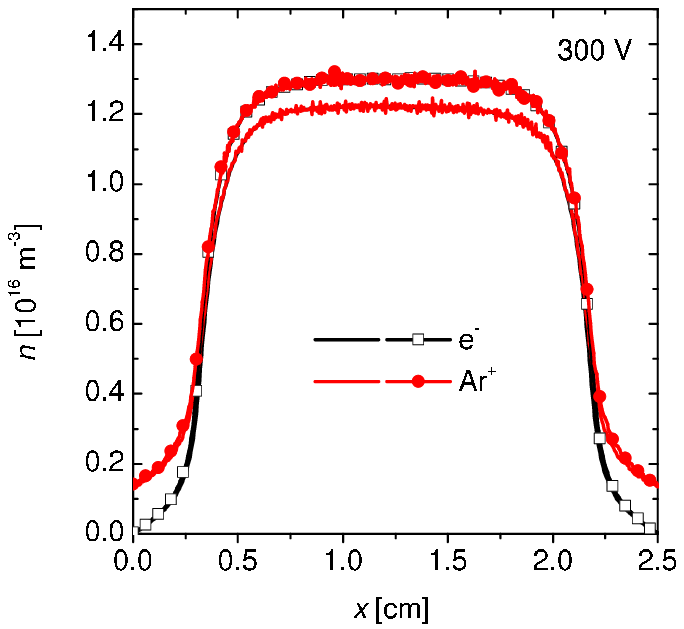}
\caption{Time-averaged charged particle density distributions obtained from PIC/MCC simulations by tracing only ions (lines - model A) and both ions and fast neutrals (lines with symbols - model D) in the simulation. Discharge conditions: 100 Pa, 13.56 MHz, 300 V voltage amplitude, 2.5 cm electrode gap, $\gamma$ = 0.}
\label{fig:densities_100Pa}
\end{center}
\end{figure}

\begin{figure}[ht!]
\begin{center}
\includegraphics[width=0.45\textwidth]{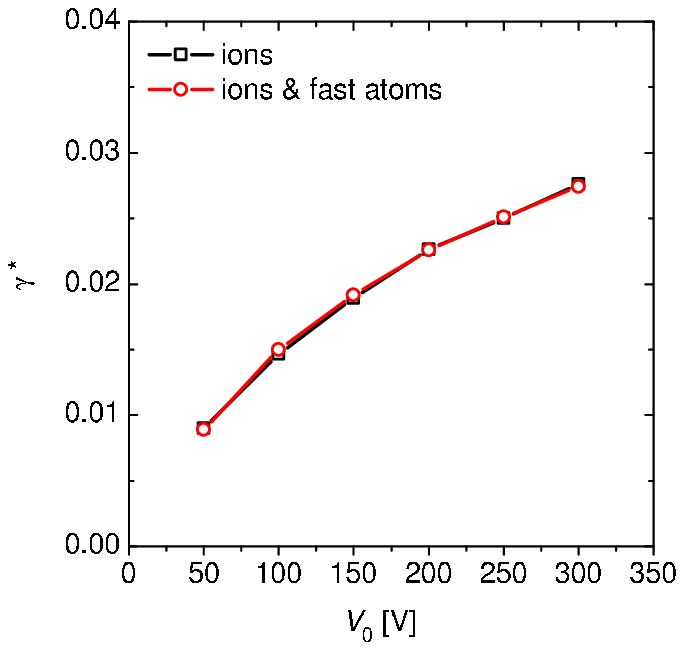}
\caption{Effective secondary electron emission coefficient as a function of the driving voltage amplitude obtained from PIC/MCC simulations using energy-dependent secondary electron emission coefficients and tracing only ions (line with open squares - model C) and both ions and fast neutrals (line with open circles - model E) in the simulations. Discharge conditions: 100 Pa, 13.56 MHz, 2.5 cm electrode gap. }
\label{fig:gamma_100Pa}
\end{center}
\end{figure}

The realistic treatment of the secondary electron emission coefficient causes an increase of all quantities depicted in figure \ref{fig:iondensity_100Pa} at high voltage amplitudes (models C and E), with an additional increase of the ion density and flux if fast atoms are considered (model E). 
Generally, the enhancement of the central ion density, the ion flux, and the mean ion energy by including secondary electron emission can again be explained by the additional ionization and the resulting smaller sheath extensions. These increases are, however, significantly larger if the secondary electron emission coefficient is set to $\gamma=0.1$ (model B), indicating an overestimation of the secondary electron yield.

\begin{figure}[ht!]
\begin{center}
\includegraphics[width=0.45\textwidth]{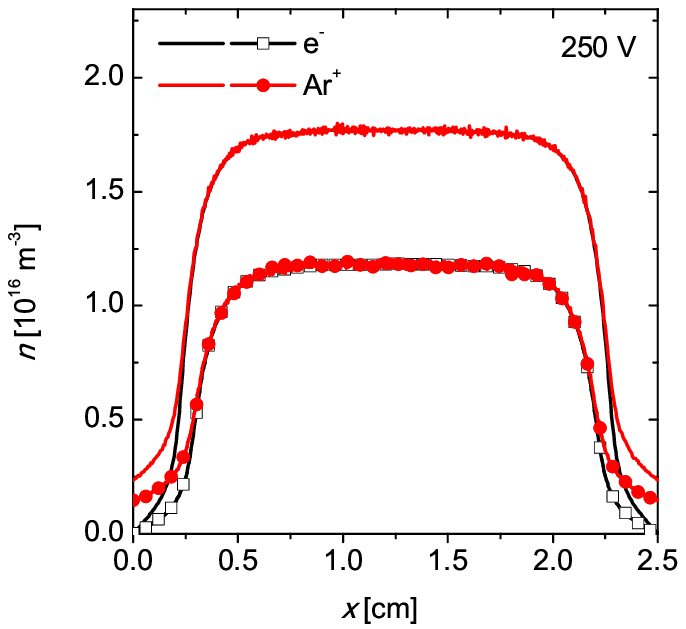}
\caption{Time-averaged charged particle density distributions obtained from PIC/MCC simulations using energy-dependent secondary electron emission coefficients and tracing both ions and fast neutrals in the model (lines with symbols -  model E), and using a constant secondary electron emission coefficient, $\gamma$ = 0.1, and tracing only ions in the model (lines - model B). Discharge conditions: 100 Pa, 13.56 MHz, 250 V voltage amplitude, 2.5 cm electrode gap.}
\label{fig:densities_100Pa_gen}
\end{center}
\end{figure}

In fact, the secondary electron emission coefficient is much smaller than 0.1, such as shown in figure \ref{fig:gamma_100Pa}: $\gamma^\ast$ is between 0.01 and 0.03 depending on the driving voltage amplitude. Such low values result from the very low energy of the heavy particles at the electrodes. Thus, these results show that the role of secondary electrons at high pressure collisional conditions is significantly less important compared to the ``classical'' assumption of $\gamma = 0.1$. Any effect of excluding (model C) or including (model E) the tracing of fast atoms on the effective secondary electron emission coefficient is negligible, because the energies of the heavy particles at the electrodes are low due to the collisional sheaths. Accordingly, the ion density and flux are only slightly larger for simulations, where fast neutrals are included in addition to ions (see figures \ref{fig:iondensity_100Pa}(a) and (b)), since only few additional secondary electrons are generated at the electrodes by fast neutrals. Subsequently, there is no noticeable effect of the additional ionization due to secondary electrons, that are emitted due to fast atom impact at the electrodes, on the sheath widths and ion dynamics in the sheath electric field, resulting in about the same  ion energy with and without tracing fast atoms (models C and E in figure \ref{fig:iondensity_100Pa}(c)).

Figure \ref{fig:densities_100Pa_gen} shows the time-averaged profiles of the electron and ion densities resulting from  model B (tracing only ions and assuming $\gamma = 0.1$) and  model E (tracing both ions as well as fast neutrals and using energy-dependent emission coefficients) at 250 V and 100 Pa. Due to the collisional sheaths and the low heavy particle energies at the electrodes, the shapes of the resulting density profiles are similar, but the assumption of $\gamma = 0.1$ results in higher central plasma densities compared to using energy-dependent emission coefficients. A comparison at higher voltage amplitudes is not possible, because the simulations of model B diverge, for the same reason as in the lower pressure case of 20 Pa. 

\begin{figure}[ht!]
\begin{center}
\includegraphics[width=0.95\textwidth]{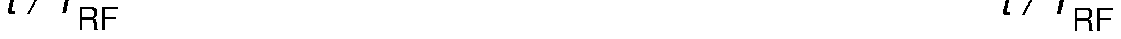}
\caption{Spatio-temporal distribution of the total ionization rate as obtained on the basis of (a)  model D (tracing fast neutrals assuming $\gamma=0$) and (b)  model E (tracing fast neutrals and calculating $\gamma^\ast$), for $p$ = 100 Pa and $V_0$ = 300 V. The color scales are logarithmic, cover two orders of magnitude, and are given in units of m$^{-3}$s$^{-1}$.}
\label{fig:totion_100Pa_gen}
\end{center}
\end{figure}

These results demonstrate that at high pressures (collisional regime) the secondary electron emission coefficients at the boundary surfaces are strongly reduced compared to those at lower pressures and less collisional regimes due to an effective reduction of the heavy particle bombardment energies by collisions inside the sheaths. Under these conditions, $\gamma = 0.1$ is unrealistically high. Our results also show that neglecting fast neutrals is justified at high pressures and low driving voltage amplitudes. 

These conclusions are further corroborated by observing the patterns of the total ionization rate, shown in figure \ref{fig:totion_100Pa_gen}. The data were computed on the basis of  model D (tracing fast neutrals and assuming $\gamma=0$) and  model E (tracing fast neutrals and calculating $\gamma^\ast$), for $p$ = 100 Pa and $V_0$ = 300 V. Figure \ref{fig:totion_100Pa_gen}(a), result of model D, indicates the full absence of ionization in the gas phase near the electrodes -- an effect that was found to exist at lower pressures. While ionization in this case occurs only near the edge of the expanding sheath, with the inclusion of secondary electrons (figure \ref{fig:totion_100Pa_gen}(b), results of model E) ionization also occurs inside the sheaths. Due to the high pressure, however, the secondary electrons accelerated here are not able to cause additional ionization deep inside the plasma bulk, in contrast to the lower pressure cases of 5 Pa and 20 Pa (see, respectively, figures \ref{fig:totion_5Pa} and \ref{fig:totion_20Pa}), where the electrons' motion is highly nonlocal. Nonetheless, inclusion of secondary electron emission is important for these conditions, too, and it is highly preferred to calculate the effective secondary electron yield via tracing of heavy particles.

\section{Conclusions}

The effects of including processes induced by fast neutrals and using realistic energy-dependent secondary electron emission coefficients due to ion and fast neutral impact at the electrodes on the spatio-temporal ionization dynamics, plasma density, ion flux, and mean ion energy obtained from PIC/MCC simulations of CCPs have been investigated systematically under conditions relevant for plasma processing applications. By studying single frequency CCPs operated in argon and driven at 13.56 MHz at 5 Pa, 20 Pa, and 100 Pa, we probe a collisionless, an intermediate, and a highly collisional regime. A systematic variation of the driving voltage amplitude is performed at each pressure and individual processes such as tracing fast neutrals, the presence of secondary electron emission from boundary surfaces, and the energy-dependence of the corresponding $\gamma$-coefficients for ions and fast neutrals are individually switched on and off to separate gas phase and surface effects of heavy particles on the discharge characteristics.

Compared to classical simulations, where $\gamma \approx 0.1$ is typically assumed, independently of the incident particle energy as well as of the surface conditions, and only ions are traced, we find significant and strong effects of tracing fast neutrals and including realistic energy-dependent $\gamma$-coefficients on the discharge characteristics. In particular, at high driving voltage amplitudes, the results of the simulations using different model assumptions deviate by up to a factor of two from one another.

At low pressures, the discharge is operated in the $\alpha$-mode. Nevertheless, the ion density and flux are strongly increased if the emission of secondary electrons is included and if fast atoms are traced. The fast atoms affect the ionization directly via gas-phase collisions and indirectly by generating electrons inside the sheaths and by increasing the effective secondary electron emission coefficient. Therefore, including fast neutrals leads to an additional ionization source in the sheaths and at the electrodes, thereby reducing the sheath widths. This results in a less collisional ion dynamics in the sheaths, leading to an increase of the ion energy at the surfaces.
These effects are even more pronounced at intermediate pressures, where the multiplication of energetic electrons inside the sheaths becomes more efficient. 
At high pressures, i.e., in a collisional regime of discharge operation, the effect of fast atoms is small, because their average energy is low. Therefore, the density, flux, and energy of ions are only weakly affected by taking fast neutrals into account.

We use electron emission yields that depend on the incident particle energy to calculate effective secondary electron emission coefficients at boundary surfaces, $\gamma^\ast$, under various discharge conditions. The results demonstrate that the secondary electron emission coefficient strongly depends on external control parameters such as the pressure and the voltage, as well as on the treatment of fast neutrals. $\gamma^\ast$ increases if the discharge voltage amplitude is increased and the pressure is reduced, respectively, as the energy of heavy particles reaching the surface increases. Furthermore, including fast atoms leads to an increase of $\gamma^\ast$ by about 50\% at low pressures. The lowest effective secondary electron emission coefficient of $\gamma^\ast \approx 0.01$ is found at a high pressure of 100 Pa and a low voltage amplitude of 100 V. This means that a constant value of $\gamma =0.1$, which is typically assumed in simulations, is too large by one order of magnitude. In general, this value can never be reached within the whole range of discharge conditions investigated here, if the role of fast neutrals is neglected. Thus, it can be concluded that using an energy-independent secondary electron emission coefficient will very likely result in an unrealistic description of the entire discharge physics.

Generally, our results show that classical PIC/MCC simulations of CCPs that do not trace fast neutrals and do not include realistic energy-dependent secondary electron emission coefficients yield unrealistic results under many discharge conditions relevant for low pressure plasma processing applications. As these simulation tools are used for process optimization -- often with important fundamental physical effects neglected --, we propose to include fast neutrals and energy-dependent surface coefficients in simulations of CCPs in order to yield more realistic results. Such more realistic simulations require only marginally longer computation times (3 \% - 5 \%) compared to simulations that trace only ions and use constant $\gamma$-coefficients under the conditions investigated here. 

\ack This work was supported by the Hungarian Scientific Research Fund through the grant OTKA K 105476.

\end{document}